\documentclass[a4paper]{article}

\usepackage{arxiv}

\usepackage[utf8]{inputenc} 
\usepackage[T1]{fontenc}    
\usepackage{hyperref} 
\usepackage{gensymb}
\usepackage{indentfirst}
\usepackage{url}            
\usepackage{booktabs}       
\usepackage{amsfonts}       
\usepackage{nicefrac}       
\usepackage{microtype}      
\usepackage{lipsum}		
\usepackage{graphicx}
\usepackage{subcaption}
\usepackage{environ}
\usepackage[round]{natbib}
\usepackage{doi}
\usepackage{comment}
\usepackage{titlesec}
\usepackage{amsmath, amsfonts, amssymb}
\usepackage{tabularx,ragged2e,booktabs,caption} 
\newcolumntype{C}[1]{>{\Centering}m{#1}}

\titlelabel{\thetitle.\quad}

\title{Effective parameterization of absorption by gaseous species and unknown UV absorber in 125-400 nm region of Venus atmosphere}

\date{\today}

\author{Boris ~Fomin\\
	Central Aerological Observatory, Moscow\\
	\texttt{b.fomin@mail.ru} \\
	
	\And
	\href{https://orcid.org/0000-0001-7117-2481}{\includegraphics[scale=0.06]{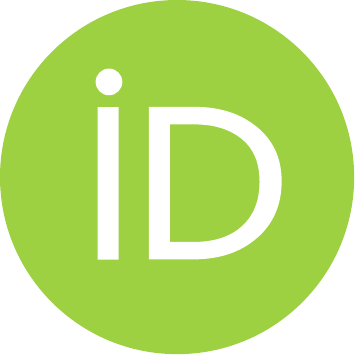}\hspace{1mm}Mikhail ~Razumovskiy\thanks{corresponding author}}\\
    
	Moscow Institute of Physics and Technology\\
	Instituskii per. 9, 141701, Moscow, Russia \\
	\texttt{razumovskii@phystech.edu} \\
}

\renewcommand{\shorttitle}{\textit{Effective parameterization of absorption in Venus Atmosphere} }

\hypersetup{
pdftitle={A template for the arxiv style},
pdfsubject={astro-ph.EP},
pdfauthor={Mikhail Razumovskiy},
pdfkeywords={Venus, radiative transfer, atmospheres, correlated-k, ultraviolet spectral region},
}

\begin{document}
\maketitle
\large
\begin{abstract}
We present an effective parameterization of molecular absorption of shortwave solar radiation in Venus atmosphere.
It is addressed to general circulation modeling to accelerate radiative transfer calculations in the
spectral interval 125 -- 400 nm (25000 -- 80000 cm$^{-1}$ ). In F-UV and M-UV regions strong
absorption of CO$_2$ ans SO$_2$ enables to parameterize gaseous absorption with only three
effective cross-sections. In N-UV region absorption of SO$_2$ and the unknown UV absorber
are parameterized with six effective cross-sections for each species. For treatment of Rayleigh
scattering and optical properties of Venus clouds eight effective spectral points are
recommended. Parameterizations were validated by the original reference line-by-line Monte-Carlo radiative transfer
model. The outcome of the validation shows the discrepancy in fluxes less than 3\%. Thus, it takes only eight-fold solution of radiative transfer equations to correctly describe solar
fluxes and heating rates in the whole ultraviolet region.
\end{abstract}

\keywords{Venus \and radiative transfer \and atmospheres \and molecular absorption \and parameterization \and unknown UV absorber}

\section{Introduction}
\label{sec:intro}

\qquad General circulation models (GCMs) require high accuracy solutions of radiative transfer equation at every grid step (spatial and temporal) in order to reproduce peculiar mesoscale structures that are notably presented in Venus atmosphere. However, it is unfeasible to include detailed line-by-line calculations to GCMs, subsequently some ways of parameterization should be involved.

\qquad By this time there is a sufficient lack of self-consistent parameterizations that whether are not too simplistic (e.g. double-gray radiative transfer or Newtonian approximation) or not too much constrained to various atmospheric parameters, specific GCM or way of solving radiative transfer equation. Thus, our motivation is to offer the fresh way to accelerate radiative calculations suitable to climate modeling. The earlier approach is based on look-up-tables of heating and cooling rates generated by \citet{Haus2017} and utilized in IPSL GCM by \citet{Garate-Lopez2018}. However, we adhere to the different concept of speeding up the radiation block of GCM.

\qquad In our approach we follow original k-distribution technique described in \citep{Fomin2005} instead of familiar correlated-k method. We construct effective cross-sections for absorption by gaseous constituents and unknown UV absorber regardless of its chemical composition. Mentioned parameterizations depend only on initial absorption cross-sections. In addition, we also suggest spectral points for calculation of Rayleigh scattering and clouds optical features (extinction coefficient and scattering indicatrix). Thus, we propose a universal tool to accelerate radiative transfer calculations. One needs to solve radiative transfer equation just eight times using our effective cross-sections to accurately reproduce fluxes and heating rates in UV region, while other blocks in atmospheric modeling scheme can be unchanged. We validate the efficiency of our parameterizations through line-by-line procedures with reference Monte-Carlo model. Construction of parameterizations of molecular absorption in other spectral regions will be the subject of future works.

\qquad The paper is organized as follows. In section \ref{sec:model} we explain the modeling part of the work: the technique for constructing the parameterizations and description of the reference Monte-Carlo model used in validation procedures. Section \ref{sec:result} refers to the dataset with parameterizations and the result of validation procedures. We provide some conclusive remarks in section \ref{sec:conclusion}.

\section{Methodology}
\label{sec:model}

\large{
\subsection{Construction of parameterizations}
}

\qquad The way we construct the parameterizations essentially corresponds to the one used in \citep{Fomin2005, Fomin2017} for shortwave region of Earth's atmosphere with some revision. In the case of Venus atmosphere in UV spectral region this method is simplified due to ignorance of absorption cross-sections temperature dependence \citep{Haus2015}. The idea of creating parameterizations is the following. Let us consider two modeling atmospheres: ``simple'' atmosphere with no scattering and no reflection from surface (A) and full-consistent Venus atmosphere (used in reference line-by-line calculations) (B). And suppose that one needs to find (``retrieve'') absorption cross-section of fixed gaseous species from given radiative fluxes obtained in the atmospheres A and B. For fixed wavelength $\lambda^*$ ``retrieval'' procedures for both atmospheres obviously will give the same value: $\sigma(\lambda^*)$, where $\sigma(\lambda)$ is the initial absorption cross-section. However, radiation calculations for atmosphere A are much faster and simpler than for atmosphere B, because they are based on the trivial Beer-Lambert law. We can follow the same procedure in simple atmosphere A but for the given spectral interval $\Delta\lambda$ and instead of monochromatic cross-sections ``retrieve'' \textit{effective cross-sections}: $\sigma_{\text{eff}}(\Delta\lambda)$. Effective cross-sections represent profile of initial absorption cross-sections averaged with solar spectrum in the interval $\Delta\lambda$. We define effective cross-section in the i-th atmospheric layer through the following formula:

\begin{equation*}
    \sigma_{\text{eff},\,i}=\frac{\text{ln}(F_{i+1}^{\,\downarrow}/F_{i}^{\,\downarrow})}{\Delta\zeta_i},
\end{equation*} 

where $F_i^{\,\downarrow}$ and $F_{i+1}^{\,\downarrow}$ are the downward fluxes at the bottom and at the top of the $i$-th discrete height layer. $\Delta\zeta_i = \zeta_{i+1}-\zeta_{i}$ \, is the number of absorber molecules along the direct solar radiation in $i$-th layer. As the absorber amount along the radiation path strongly correlates with the zenith angle, it is more useful to introduce dependence of the effective cross-sections from the value $\zeta$, instead of height $h$. Thus defined values $\sigma_{\text{eff},\,i}$ can be used in one-time solution of radiative transfer equation to find radiative fluxes and heating rates in the given spectral interval $\Delta\lambda$.

\qquad More wider the interval $\Delta\lambda$ -- more bigger the difference between radiative fluxes and heating rates calculated for simple atmosphere A and ``real'' atmosphere B. Nevertheless, we found out that spectral intervals where effective cross-sections obtained in simple atmosphere A yield good accuracy, appear to be very large. Thus, we divide 125-400 nm region into eight  \textit{k-intervals}, where effective cross-sections yield good accuracy in fluxes and heating rates. For providing stable justification of our approach, we validate our parameterizations with reference Monte Carlo model. During the validation procedures for each k-interval we compare radiative fluxes and heating rates calculated in two ways: as a result of one-time solution of radiative transfer equation using effective cross-section for this k-interval and as a result of line-by-line calculations in this k-interval with reference Monte Carlo model.

\qquad After thorough manual inspection of the whole UV spectral interval, we divided it into 8 k-intervals to calculate effective cross-sections for each of them. In each k-interval we selected wavelength points for calculation of Rayleigh scattering in order to obtain best fits with reference simulations. For convenience, optical features of the Venus clouds (extinction coefficient and scattering indicatrix) could be taken at the same point since they don't significantly depend on wavelength in UV region. We also specify the incident solar flux at the top of the atmosphere (TOA) for each k-interval as this value relates to the way of choosing solar spectrum data. All detailed information about the k-intervals is summarized in Table \ref{tab:k-intervals}.

\bigskip
\begin{minipage}{\linewidth}
\centering
\captionof{table}{Detailed information about k-intervals for accelerating radiative calculations in UV region} 
\label{tab:k-intervals} 
\begin{tabular}{ C{1.0in} C{1.4in} C{1.2in} C{0.8in} C{1.4in}}\toprule[1.5pt]
\bf k-interval & \bf Spectral region, $\bf 10^4\text{cm}^{-1}$ & \bf Superior absorbers & \bf ${\bf \nu_R}$, $\bf \text{cm}^{-1}$ & \bf Incident solar flux, W/m$^{\bf 2}$ \\\midrule
F-UV & 5 - 8 & CO$_2$ & 66000 & 0.20\\  
		M-UV & 3.(3) -- 5 & SO$_2$, CO$_2$ & 36000 & 27.74\\ 
		
		N-UVa & 3.1 -- 3.(3) & SO$_2$, UVA & 32000 & 28.19\\ 
		
		N-UVb & 3.0 -- 3.1 & SO$_2$, UVA & 30000 & 19.91\\ 
		
		N-UVc & 2.9 -- 3.0 & UVA, SO$_2$ & 30000 & 20.20\\ 
		
		N-UVd & 2.7 -- 2.9 & UVA, SO$_2$ & 28000 & 52.40\\ 
	
		N-UVe & 2.55 - 2.7 & UVA, SO$_2$ & 26250 & 17.64\\ 
		
		N-UVf & 2.5 - 2.55 & UVA, SO$_2$ & 25250 & 47.27\\
		
\bottomrule[1.25pt]
\end {tabular}\par
\bigskip
\normalsize{ 
\begin{itemize}
    \item $\nu_R$ -- is the recommended wavenumber for calculation of Rayleigh scattering and clouds optical features during one-time radiative transfer simulation
    \item UVA relates to the unknown ultraviolet absorber
    \item The left border of M-UV region was chosen for convenience (300 nm)
    \item More dominant absorber goes first. Other active absorbers can be neglected while considering the whole UV-region (more in Section \ref{subsec:validation}).
\end{itemize}
}
\end{minipage}

\subsection{Monte-Carlo reference model}
    
    \qquad Our reference Monte-Carlo model was previously successfully exploited for Earth' conditions in UV region by \citet{Sukhodolov2016}. It is a high-resolution (0.25 cm$^{-1}$) plane-parallel short-wave version of FLBLM model \citep{Forster2011, Fomin2012}. For the present studies it was adopted to the Venus conditions according to the baseline work by \citet{Haus2015}, in which authors performed line-by-line calculations along with sensitivity analysis to various spectroscopic and atmospheric parameters from 125 nm to 1000 $\mu$m. Hereafter we list the key parameters of the atmospheric model. 

\subsubsection{UV absorption cross-sections}    
\begin{figure}
  \centering
 \includegraphics[scale=1.0]{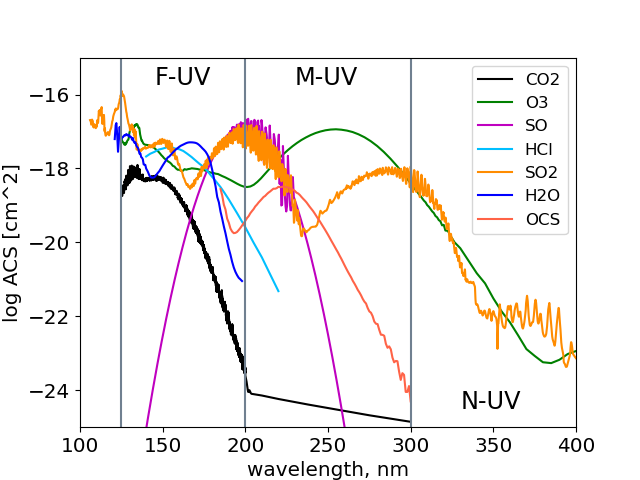}
  \caption{Absorption cross-sections of gaseous constituents in Venus atmosphere in UV region}
  \label{fig:photoabsorption}
\end{figure}

    \qquad Photoabsorption spectra of gaseous components in the atmosphere are relatively smooth in comparison with visible and IR-spectra. This fact is largely determined by the nature of absorption of UV rays: it drives the course of the chemical reaction of photodissociation. Thus, to incorporate UV spectroscopy into simulations, there is no need to trace the contours of individual lines. Relevant sources on absorption cross-sections of atmospheric gaseous constituents are summarized in Table \ref{tab:photoabsorption}. Note that in some experiments extinction coefficients or absorption coefficients were measured, which are not so convenient in our studies. We refer to MPI-Mainz spectral atlas \citep{Keller-Rudek2013} to gain datasets in unique format: absorption cross-section in cm$^2$/molecule from wavelength in nm. Figure \ref{fig:photoabsorption} shows absorption cross-sections as functions from wavelength for all gaseous constituents which are considered to be active in UV-region. The strongest absorption for all gaseous components is manifested in the far ultraviolet in 125-200 nm, where the intensity of solar radiation is insignificant. In present paper cross-sections of carbon dioxide are represented as a combination of data from \citet{Shemansky1972}, \citet{Parkinson2003}, \citet{Yoshino1996} in different spectral intervals. As \citet{Haus2015} we also neglected the temperature dependence of the cross-sections for all gases. Nonetheless, \citet{Marcq2019} consider temperature dependent absorption cross-sections for CO$_2$ and SO$_2$. We address this possible issue in Section \ref{subsec:validation}.

    \begin{table}[!htbp]
	\caption{\bf Sources of wavelength-dependent absorption cross-sections}
	\centering
	\begin{tabular}{llll}
		Species & Spectral region & Source paper & Temperature\\
		\midrule
		CO$_2$ & 125 -- 163 nm & \citep{Yoshino1996} & 295K\\  
		CO$_2$ & 163 -- 200 nm & \citep{Parkinson2003} & 295K\\ 
		CO$_2$ & 200 -- 300 nm & \citep{Shemansky1972} & 298K \\   
		SO$_2$ & 125 -- 400 nm & \citep{Manatt1993} & 293K\\ 
		SO & 125 -- 330 nm & \citep{Phillips1981, Belyaev2012} & 293K\\
		O$_3$ & 125 -- 186 nm & \citep{Mason1996} & 298K\\
		O$_3$ & 186 -- 400 nm & \citep{JPL2011} & 293-298K\\
		H$_2$O & 125 -- 198 nm & \citep{JPL2011} & 298K   \\
		HCl & 140 -- 220 nm & \citep{Inn1975} & 298K \\
		OCS & 185 -- 300 nm & \citep{Molina1981} & 295K \\
		\bottomrule
	\end{tabular}
	\label{tab:photoabsorption}
\end{table}
    
\subsubsection{Unknown UV absorber}    
        \qquad The unknown UV absorber (UVA) is considered to be an additional source of radiation absorption in the spectral range 320-400 nm \citep{Pollack1980}. We include the UVA to our atmospheric model according to one of the latest parameterizations by \citet{Haus2015} (we use in simulations both high-altitude and low-altitude models, fig. 18) in contrast to \citet{Crisp1986, Marcq2019} who effectively considered it by depressing single scattering albedo of cloud mode 1. \citet{Haus2015} extracted (``retrieved'') the absorption cross-sections and concentration in such a way that the simulation results were in adequate agreement with the experimental Bond albedo data by \citet{Moroz1981}. This approach allows to treat unknown UV absorber as another species with its own concentration profile and absorption cross-section without specifying a concrete molecular formula. Though nominal high-altitude model given by \citet{Haus2015} features only 10 particles per cm$^3$ in 58-72 km region, huge absorption cross-sections conduce effective optical depths of UV absorber comparable to corresponding ones of SO$_2$. The active debate about the chemical composition of UVA is still going and the cycle of sulfur species in the upper clouds layer is not yet fully understood. Because the frontier OSSO parameterization by \citet{Frandsen2016} didn't yield realistic radiance factors in radiation scheme in \citep{Marcq2019}, we decided to keep ``retrieved'' cross-section by \citet{Haus2015} regardless of chemical composition.
    
\subsubsection{Other parameters}
\label{sec:other_param}
    \qquad According to \citet{Haus2015, Marcq2019}, CO$_2$, SO$_2$, SO, H$_2$O, OCS, HCl and O$_3$ are the only active gases in UV-region which may possibly affect the outcome of radiation simulations. CO$_2$ abundance is assumed to be uniform in height and equals to 0.965. Due to the strong variability of SO$_2$ concentrations we incorporate two vertical profiles in radiative transfer simulations. The first one is taken from \citet{Haus2015} fig. 5, that corresponds to 150 ppmv below the clouds, 15\% larger than recommended by \citet{Bezard1993}. The second profiles features abundance of SO$_2$ below the clouds in according with \citet{Vandaele2017}, fig 1. Meanwhile, concentrations above the clouds in this profile is taken from theoretical models of \citet{Zhang2012}. SO:SO$_2$ ratio is set to be constant and equal to 10\% \citep{Marcq2019, Marcq2020}. The volume mixing ratio of water vapor is assumed to be 32.5 ppmv below 50 km \citep{Marcq2009,Arney2014}  and 3 ppmv above 70 km \citep{Cottini2012, Fedorova2016}. Recently discovered ozone is accounted in radiative transfer simulations via profile from theoretical model by \citet{Krasnopolsky2013}, which features typical abundance of 10 ppbv at 90 km. OCS concentrations are significant only below clouds, while HCl volume mixing ratio is set to be 0.5 ppmv for all heights. However, during validation we show that most radiation redistribution occurs only up above the clouds layer, so the influence of H$_2$O, OCS and HCl is negligibly small. 
    
    \qquad Strength of the Rayleigh scattering is proportional to fourth power of wavenumber. Thus, it is assumed to be a crucial ingredient in radiation simulations in UV region. We take Rayleigh scattering into consideration in the way similar that \citet{Haus2015} do. They made slight correction to formula of \citet{Hansen1974} for pure CO2 atmosphere, changing the surface pressure. We use the same value of $p_0=92.1$ bar in agreement with VIRA model. Despite Rayleigh scattering optical depth surges while going to the depth of the atmosphere, it will be shown that in UV region taking this scattering into account is minor compared to gaseous absorption. For example in F-UV region it can be completely neglected as it manifests itself in heights where nearly all the radiation is already absorbed.
    
    \qquad Solar spectrum in the UV region may vary slightly depending on solar activity. We utilized the COSI spectrum from \citet{Shapiro2010, Shapiro2011}, which was recommended in \citep{Sukhodolov2016}. As shown by numerical experiments, usage of a various spectra does not significantly affect the resulting parameterizations of gaseous absorption. It is only required to change accordingly the solar fluxes in each k-interval falling on the upper boundary of the atmosphere. 
    
    \qquad We used fixed VIRA \citep{Zasova2006} thermal profile for equatorial region on the night side from 0 to 150 km in our modeling purposes. With this approximation we abide by \citet{Marcq2019} due to above mentioned assumption of temperature independence of absorption cross-sections. So, the present parameterization could not be affected by applying different temperature profiles. However the heating rates depend on pressure, we expect that they predominantly are influenced by absorption features and incident solar flux.
    
    \qquad Our simulating scheme uses recent cloud model that consists of four modes of 75\% H$_2$SO$_4$ spherical aerosols \citep{Pollack1993, Zasova2007, Haus2013}. To find microphysical parameters we apply Mie theory code \citep{Fomin1998}. All four modes are log-normally distributed with parameters of modal radii of 0.3, 1.0, 1.4, 3.65 $\mu$m and unitless dispersions of 1.56, 1.29, 1.23, 1.28, respectively \citep{Pollack1993, Haus2013}. Refractive indices data were taken from \citet{Palmer1975}. Scattering properties do not significantly alter with wavelength in UV region. Thus, any other clouds parameters may be easily inserted in atmospheric model, as our gaseous absorption parameterization is obtained regardless of clouds features.
    
\section{Result and validation with the reference model}
\label{sec:result}

\qquad Original data of this study (mainly including resulting effective cross-sections) are available at Mendeley Data (\href{https://dx.doi.org/10.17632/97rskzggbj.1}{https://dx.doi.org/10.17632/97rskzggbj.1})

\subsection{Effective cross-sections}

\begin{figure}
    \centering
        \includegraphics[width=0.48\textwidth]{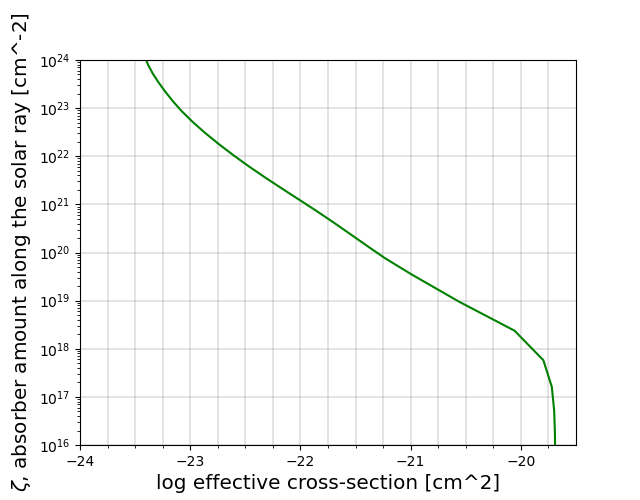}
        \hfill
        \includegraphics[width=0.48\textwidth]{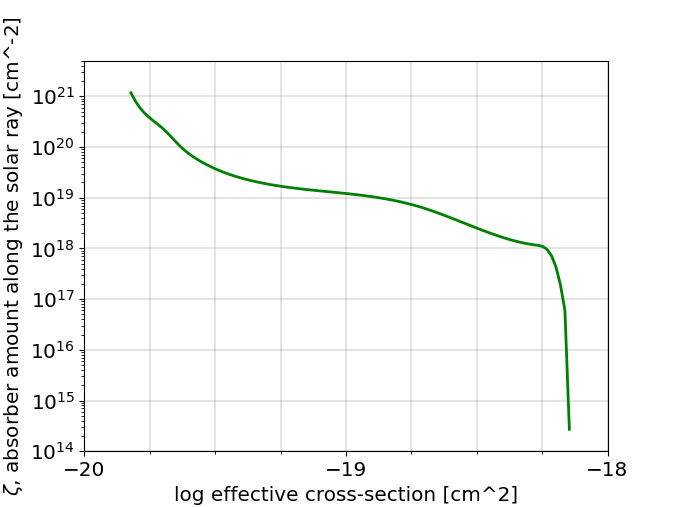}
        \caption{CO$_2$ (left) and SO$_2$ (right) effective cross-sections in F-UV and M-UV k-intervals respectively}
        \label{fig:cross_sec}
\end{figure}

\qquad In this section we present some information about the resulting parameterizations. In all tables $\sigma$ is an effective cross-section (in cm$^2$ per one molecule) and $\zeta$ is the absorber amount along the direct solar radiation from TOA to the given point (in molecules per cm$^2$). Tables are given for eight k-intervals defined in Table \ref{tab:k-intervals}. Fig.\ref{fig:cross_sec} shows effective cross-sections for CO$_2$ and SO$_2$ in F-UV and M-UV k-intervals respectively. Obtained effective cross-sections are literally the result of averaging the initial absorption cross-sections over the solar spectrum reached the current atmospheric level. At the upper levels, solar radiation is still weakly transformed and the averaging is actually carried out over the initial solar spectrum. While penetrating in the atmosphere, the solar spectrum changes so that the radiation remains only in those spectral regions where the absorption is weak. This explains the decrease in effective cross-sections with increasing absorber amount.

\subsection{Validation}
\label{subsec:validation}
\begin{figure}
    \centering
        \includegraphics[width=0.48\textwidth]{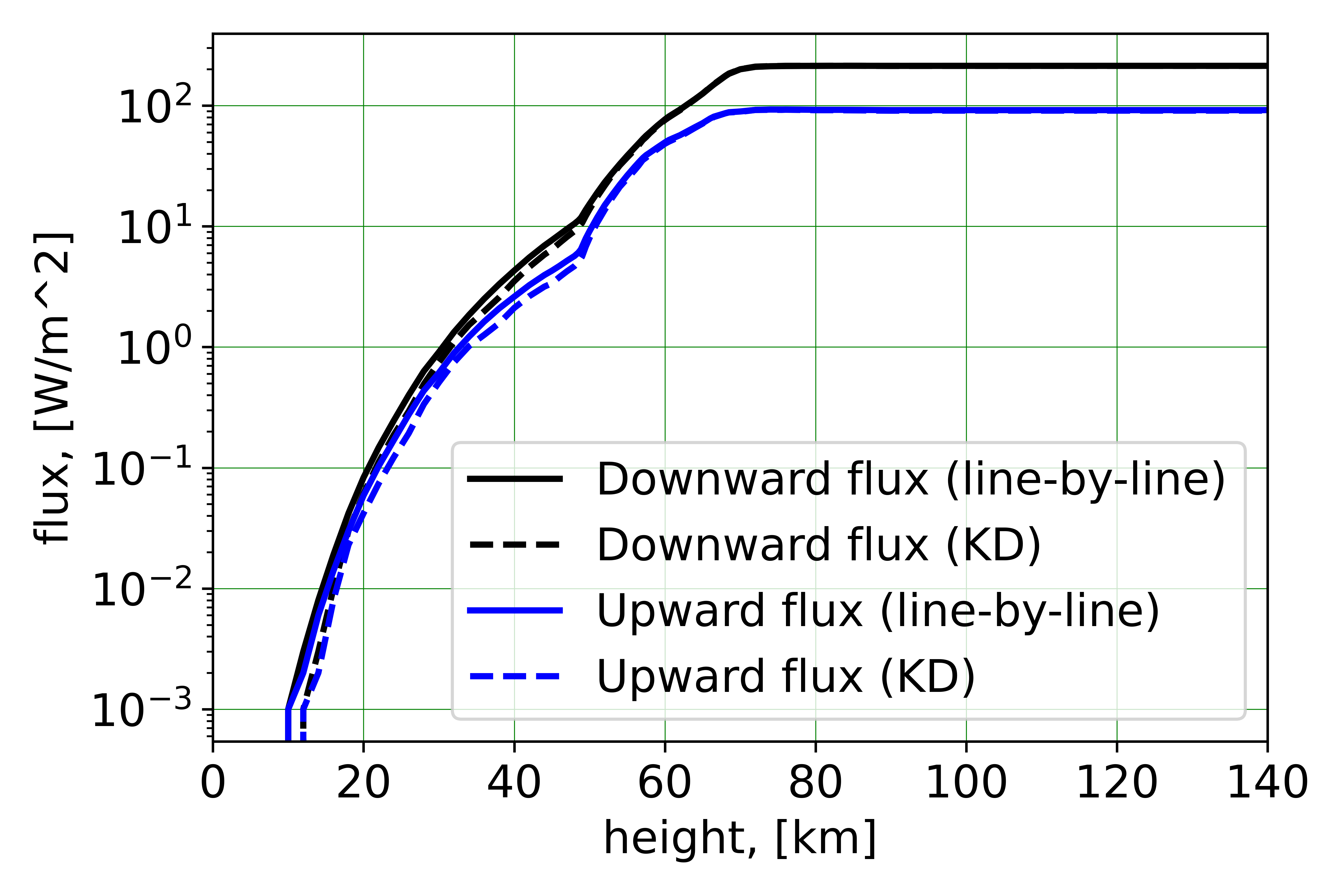}
        \hfill
        \includegraphics[width=0.48\textwidth]{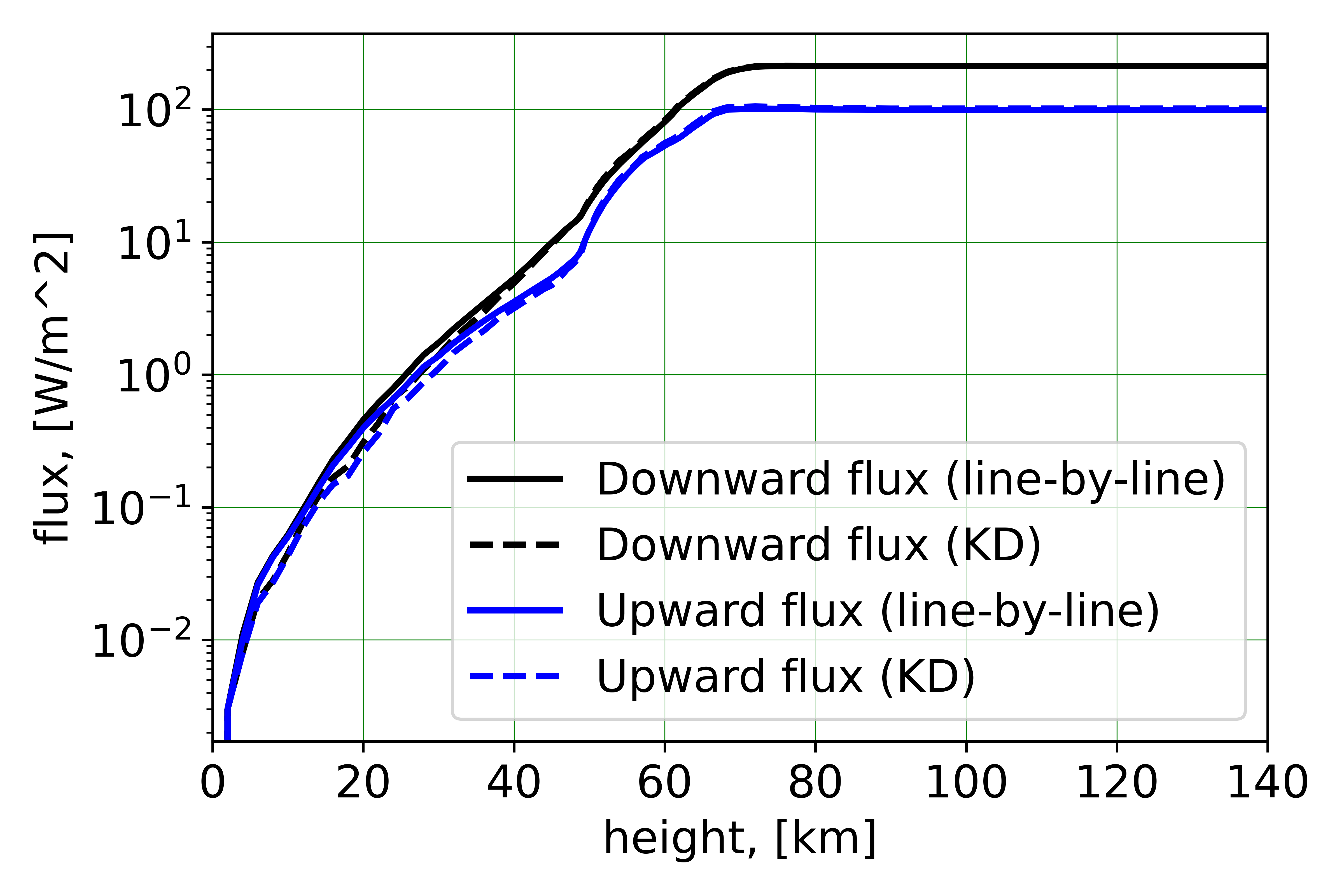}
        \hfill
        \caption{Downward and upward fluxes obtained from Monte-Carlo line-by-line simulations and from fast simulations with employing k-distribution (KD) technique with effective cross-sections. Profiles of SO$_2$ are taken from \citet{Haus2015} (left) and from \citet{Zhang2012,Vandaele2017} (right). Spectral interval 125 -- 400 nm.}
        \label{fig:validation}
\end{figure}

\begin{figure}
    \centering
        \includegraphics[width=0.48\textwidth]{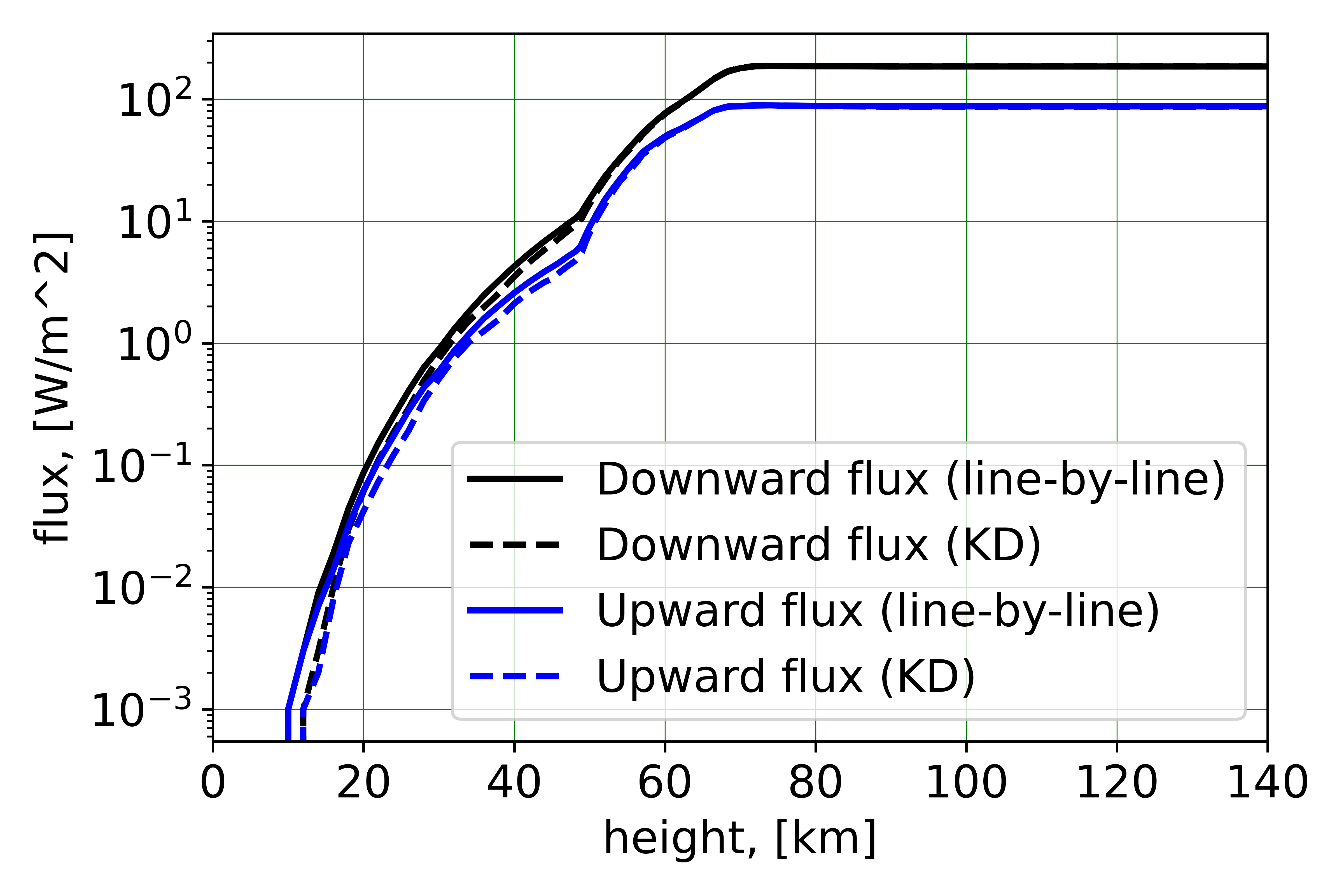}
        \hfill
        \includegraphics[width=0.48\textwidth]{XXXold_ET-KD_0deg.25-33_fig.png}
        \hfill
        \caption{Downward and upward fluxes obtained from Monte-Carlo line-by-line simulations and from fast simulations with employing k-distribution (KD) technique with effective cross-sections. Profiles of unknown ultraviolet absorber are: high-altitude model (left) and low-altitude model (right) by \citet{Haus2015}. Spectral interval 300 -- 400 nm.}
        \label{fig:validation2}
\end{figure}

\begin{figure}[htb]
  \centering
  \includegraphics[scale=0.65]{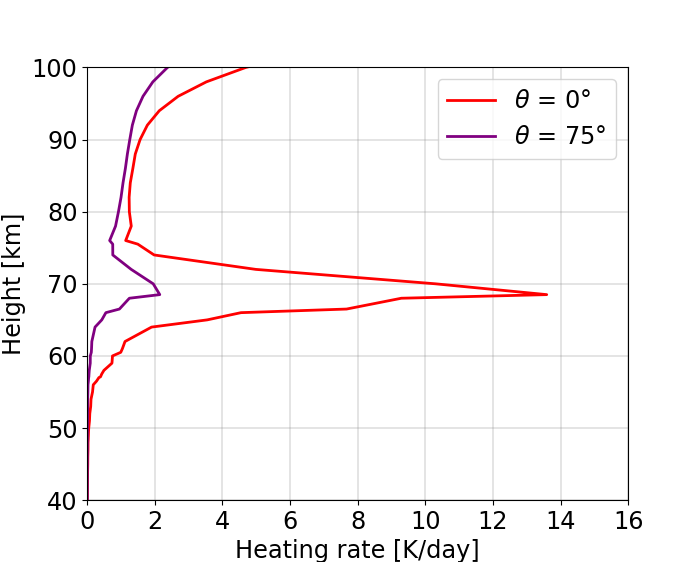}
  \caption{Solar heating rates for zenith angles $\theta=0\,\degree$\, and $\theta=75\,\degree$\, for Venus middle atmosphere. SO$_2$ and UVA profiles are taken from \citet{Haus2015}. Spectral interval 125 -- 400 nm. Dashed lines for the fast (KD) calculations are almost invisible due to good fit.}
  \label{fig:valid_heating}
\end{figure}

\qquad Validation of the presented parameterizations was performed by comparison of downward and upward solar fluxes and heating rates obtained through fast and reference simulations. Figs. \ref{fig:validation} and \ref{fig:validation2} show upward and downward fluxes for different atmospheric profiles for 0\degree \, zenith angle. Fig. \ref{fig:validation} displays results of calculations with two different SO$_2$ profiles (see \ref{sec:other_param}), keeping UVA profile fixed (high-altitude model by \citet{Haus2015}). On fig. \ref{fig:validation2} we present results of calculations with high-altitude and low altitude models by \citet{Haus2015}, keeping fixed SO$_2$ profile from the latter study. For verification the sustainability of our approach we also checked the parameterizations for two solar zenith angles: 0\degree  \,and\, 75\degree\, (see fig. \ref{fig:valid_heating}). Featured heights lie in range below 100 km. Validation displays good fits, resulting with relative discrepancies of less than 3\% for fluxes and less than 1\% for heating rates. During this procedure we also could analyze the contribution of each absorber. Atmospheric O$_3$ and SO play minor role in UV interval, less than 1 K/day contribution to the heating rate. However, we added cross-sections of this species to the dataset. Thus, we state that only CO$_2$, SO$_2$ and unknown UV absorber are relevant in terms of examination the integrated UV fluxes and heating rates.

 \qquad We investigated temperature dependence of absorption cross-sections of CO$_2$ in F-UV region taking data of Parkinson for 195 and 295 K \citep{Yoshino1996, Parkinson2003} and linearly extrapolating it between those temperatures. Resulting heating rates discrepancies appeared to be less than 1\%. We expect that for climate modeling purposes the temperature dependence could be ignored as the relevant atmosphere is from 50 -- 100 km where cross-sections doesn't change dramatically. However, we do not dismiss the interest that future research should be conducted on this issue.

\section{Conclusion}
\label{sec:conclusion}
\qquad In this work we present parameterizations of absorption by gaseous species and unknown UV absorber in 25000 - 80000 cm$^{-1}$ spectral interval. Such parameterizations are determined only by wavelength-dependent absorption cross-sections. Thus, it easily can be inserted in any radiative transfer scheme of any GCM. The result was validated with the help of original high-resolution line-by-line Monte-Carlo radiative transfer model. We expect that such parameterizations can facilitate Venus climate modeling studies. Future works will be dedicated to creating effective parameterizations of molecular and clouds absorption in visible and infrared spectral interval.
\newpage

\bibliography{references} 

\end{document}